\documentclass[10pt,a4paper,final]{article}

\usepackage[left=4.00cm,right=4.00cm,top=3.50cm,bottom=3.50cm]{geometry}

\usepackage[english]{babel}

\usepackage[hidelinks]{hyperref}
\usepackage{amssymb}
\usepackage{amsbsy}
\usepackage{amsfonts}
\usepackage{amsmath}
\usepackage{multirow}
\usepackage[labelformat=simple]{subcaption}

\usepackage{graphicx}

\usepackage{array}
\usepackage{url}

\DeclareMathOperator*{\argmax}{arg\,max}
\DeclareMathOperator*{\argmin}{arg\,min}
\DeclareMathOperator{\diag}{diag}

\usepackage[font={small}]{caption}

\begin{document}
\title{\vspace*{-6.00mm}\bfseries Markov Chain Modeling and Simulation\\of Breathing Patterns\footnotemark[1]\footnotetext[1]{Submitted for publication.}}

\author{Davide~Alinovi\textsuperscript{1}, Gianluigi~Ferrari\textsuperscript{1}, Francesco~Pisani\textsuperscript{2}\\and Riccardo~Raheli\textsuperscript{1}}

\date{}

\maketitle

\begin{center}
\vspace*{-4.0mm}
{\small
\textsuperscript{1}Department of Information Engineering, University of Parma, Italy\\
\textsuperscript{2}Department of Neuroscience, University of Parma, Italy\\
E-mail:~alinovi@tlc.unipr.it, \{gianluigi.ferrari, francesco.pisani, riccardo.raheli\}@unipr.it
\vspace{1.0mm}}
\end{center}

\section*{Abstract}
The lack of large video databases obtained from real patients with respiratory disorders makes the design and optimization of video-based monitoring systems quite critical.
The purpose of this study is the development of suitable models and simulators of breathing behaviors and disorders, such as respiratory pauses and apneas, in order to allow efficient design and test of video-based monitoring systems.
More precisely, a novel Continuous-Time Markov Chain (CTMC) statistical model of breathing patterns is presented.
The Respiratory Rate (RR) pattern, estimated by measured vital signs of hospital-monitored patients, is approximated as a CTMC, whose states and parameters are selected through an appropriate statistical analysis. Then, two simulators, {software-} and {hardware-based}, are proposed.
After validation of the CTMC model, the proposed simulators are tested with previously developed video-based algorithms for the estimation of the RR and the detection of apnea events.
Examples of application to assess the performance of systems for video-based RR estimation and apnea detection are presented.
The results, in terms of Kullback-Leibler divergence, show that realistic breathing patterns, including specific respiratory disorders, can be accurately described by the proposed model; moreover, the simulators are able to reproduce practical breathing patterns for video analysis.
The presented CTMC statistical model can be strategic to describe realistic breathing patterns and devise simulators useful to develop and test novel and effective video processing-based monitoring systems.

\subsubsection*{Keywords}
Breathing modeling, video simulation, apnea simulation, respiratory rate analysis, apnea detection.

\section{Introduction}
\label{sec:Intro}
The Respiratory Rate (RR) is a fundamental vital sign to assess the health condition of a patient: for this reason, it may be important to monitor this parameter continuously in several clinical scenarios.
Anomalous trends or values of this parameter can be the sign of a respiratory disease, such as Biot's breathing~\cite{WYKA2011}, Kussmaul's breathing~\cite{WYKA2011}, Cheyne-Stokes's breathing~\cite{ChStResp90} or Ondine's curse~\cite{Healy}, also referred to as Congenital Central Hypoventilation Syndrome (CCHS).
More generally, RR abnormal behaviors can be a sign of critical medical conditions.
In some cases, they can be an indicator of a potentially deadly event, such as an apnea, which can be defined as a persistent absence of breath or a too low RR. Hence, it is very important to promptly detect these events, which may be occasionally fatal if untreated.
Current measurement systems of the RR, also used for apnea detection, are based on polysomnographic devices, which are composed of several sensors.
Nevertheless, these systems have some drawbacks: (i) they are expensive and can be used in hospital environments only, (ii) they require specialized staff and (iii) they are moderately invasive due to wired sensors, especially for newborns.

Alternative monitoring systems could yield significant improvements in the welfare of the patients. Hence, non-invasive, low-cost, wireless monitoring and diagnostic systems are under development.
Thanks to the recent miniaturization of sensors, wearable health monitoring systems can help to monitor a patient continuously.
In~\cite{WearSurvey10}, modern techniques for the extraction of physiological signals, also related to respiration, are presented.
They rely on low-cost technologies and can be a replacement for many sensors used in the clinical environment, despite the fact that they require a ``direct connection'' to the patient.
Contactless RR long-term monitoring, based on the use of ultrasonic sensors for precise distance measurements~\cite{MiKiShHyLeLe10} or the received signal strength in a wireless network~\cite{Patwari14}, were also developed.
Among contactless monitoring systems, properly designed video-processing algorithms are of significant interest.
In~\cite{Embedded,Contactless,RealTime}, contactless monitoring systems are proposed: the first system is embedded in a board with multiple cameras~\cite{Embedded}, the second one analyzes respiratory movements, but does not include automatic RR estimation~\cite{Contactless} and the last one makes use of infrared cameras~\cite{RealTime}.
Some recent innovative video-based systems for RR measurement and apnea detection are based on advanced video-processing algorithms to enhance small breathing motion, improve apnea event detection, and refine RR estimation~\cite{BIOMS14,MEMEA15}.

A difficulty in the design of video processing-based algorithms is the lack of large databases of relevant video recordings properly matched with reliable medical data, due to the rarity of CCHS and severe apnea events, especially in full-term newborns.
For this reason, the development of a statistical model of RR patterns, including the occurrence of apnea events, is of significant interest.
Such a model can be very useful in order to devise realistic simulators and create a large set of video recordings which allow a more efficient design of automatic RR estimation and apnea detection systems.

In the literature, some physically-based anatomical simulators have been presented.
In~\cite{CVRMed}, a hardware system to handle bio-mechanical movements and simulate an anatomical and functional model of the evolution of the human trunk structures during respiration is proposed.
In~\cite{BreathEasy}, a system of rigid and deformable parts, which simulates the biological function of respiration for computer animation, is presented.

In this paper, a statistical model, based on a Continuous-Time Markov Chain (CTMC), aimed at simulating the main features of a realistic RR pattern, is derived from medical data.
The model parameters are extracted by an inference system for continuous-time Markov random processes.
Afterward, the described model is used as background for the definition of two simulators.
A \emph{software-based} simulator, able to directly manipulate video recordings of regularly breathing patients in order to introduce artificial breathing disorders, is first presented.
A \emph{hardware-based} simulator is also developed: it exploits a manikin equipped with a moving chest to physically reproduce possible breathing disorders according to the proposed statistical model.
The developed simulators are then used to test video processing-based algorithms for RR monitoring.
This paper expands upon preliminary work appeared in~\cite{MUSTEH15}, where a two-state model of apnea episodes was proposed.

The rest of the paper is organized as follows. In Section~\ref{sec:RRStatModel}, the CTMC-based RR statistical model is presented. Section~\ref{sec:Simulators} describes the two developed simulators, software- and hardware-based. Section~\ref{sec:ApplRes} addresses the validation of the statistical model and the resulting simulators on the basis of previously developed video-based monitoring algorithms. Finally, in Section~\ref{sec:Conclusion} conclusions are drawn.

\section{Respiratory Rate Statistical Model}
\label{sec:RRStatModel}
The RR is commonly defined as the number of breathing cycles per time unit, typically expressed in breaths per minute~[bpm] or, alternatively, in cycles per second~[Hz], where a breathing cycle consists of a complete sequence of inhalation and exhalation movements.
The RR changes over time, depending on physical activity and health conditions.
Normally, the RR of a patient at rest is age-dependent and typically ranges from $30$~bpm to $60$~bpm (equivalent to $0.5$--$1.0$~Hz) for newborns and from $12$~bpm to $20$~bpm for adults (equivalent to $0.2$--$0.333$~Hz)~\cite{WYKA2011}.

In order to devise a simple model of the RR pattern, it is useful to introduce a finite set of states $\mathcal{S} = \left\lbrace S_{0}, S_{1}, \ldots ,S_{N-1} \right\rbrace$.
State $S_{n}$, with $n \in \left\lbrace 0,1,\ldots,N-1 \right\rbrace $, describes breathing with a RR denoted as $\varrho_{n} \in \mathbb{R}^{+}$.
Occurrence of respiratory pauses or apnea events and large random movements of the patient body are also considered.
The statistical model of the RR pattern can encompass all the following conditions.
\begin{itemize}
\item If the patient is regularly breathing, i.e. he/she is not suffering from apnea events and no large random body movements appear, the states $\left\lbrace S_{0},S_{1},\ldots,S_{N-1} \right\rbrace$ are used to describe regular RRs, characterized by values $\left\lbrace \varrho_{n} \right\rbrace_{n=0}^{N-1}$ with $\varrho_{n} \in \left[ R_{L},R_{H} \right]$, where $R_{L} > 0$, $R_{H} > R_{L}$ denote lowest and highest admissible RRs, respectively.
\item If the patient is affected by respiratory pauses/apneas, then the state $S_{0}$ is reserved to represent this condition, so that $\varrho_{0}$ is formally set to~$0$, to describe absence of breathing and states $\left\lbrace S_{1},S_{2},\ldots,S_{N-1} \right\rbrace$ are considered for regular breathing.
\item If the patient is subject to large random body movements, during which the RR is undetectable, the state $S_{N-1}$ is reserved to represent this condition. The RR $\varrho_{N-1}$ is set to an arbitrary value $R_{M}$ much larger than the physically acceptable ones: more precisely, $\varrho_{N-1}$ is set to $R_{M} \gg R_{H}$. States $\left\lbrace S_{0},S_{1},\ldots,S_{N-2} \right\rbrace$ are still used to represents regular RRs.
\item If the patient is both suffering from respiratory pauses/apneas and subject to large random body movements, the states $S_{0}$ (with $\varrho_{0} = 0$) and $S_{N-1}$ (with $\varrho_{N-1} = R_{M} \gg R_{H}$) are reserved for absence of breathing and random movements, respectively. The remaining states $\left\lbrace S_{1},S_{2},\ldots,S_{N-2} \right\rbrace$ are used to describe regular breathing.
\end{itemize}
The following ordering is assumed: $\varrho_{0} < \varrho_{1} < \cdots < \varrho_{N-1}$.
Since the RR is inherently continuous-valued, each state represents an approximation of the real RR.
Therefore, the set $\mathcal{S}$ represents a finite state model of a discrete-valued process approximating the overall RR pattern.
The larger the number $N$ of states, the better the approximation at the cost of a higher modeling complexity.

\begin{figure}[t]
	\centering
	\includegraphics[width=0.66\textwidth]{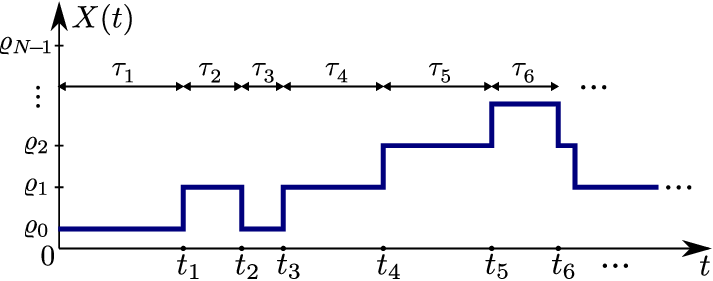}
	\caption{An example of RR pattern modeled by the finite set $\mathcal{S}$, showing sojourn times and jump times.}
\label{fig:1}
\end{figure}

According to the above statistical model, the RR process, denoted as $X(t)$, is defined as a continuous-time process with state space $\mathcal{S}$.
The time intervals during which the patient is breathing with rate $\varrho_{n}$ or is subject to apnea/respiratory pause or large body movements, namely the sojourn times in the corresponding state $S_{n}$, can be modeled as random variables and the introduced random process $X(t)$ can be generally described as a Markov process.
Ignoring the influence of other vital signs which can modify the RR of a patient over time, such as the heart rate or the oxygen saturation in the blood, the RR pattern cannot be predicted.
To derive a model that approximates this stochastic behavior, let us introduce the random variable $\tau_{\ell}$, which specifies the $\ell \text{-th}$ sojourn time, where $\ell \in \mathbb{N}^{+}$ is an index that counts the number of state changes.
Jump times can be expressed, in terms of sojourn times, as
\begin{equation}
	t_{\ell} = \sum_{q=1}^{\ell}\tau_{q}\text{.}
\label{eq:TotalTimes}
\end{equation}
In Figure~\ref{fig:1}, a graphical example of the modeled finite-state RR process $X(t)$ is shown, with highlighted sojourn times and change of state instants.

\begin{figure}[b]
	\centering
	\includegraphics[width=0.55\textwidth]{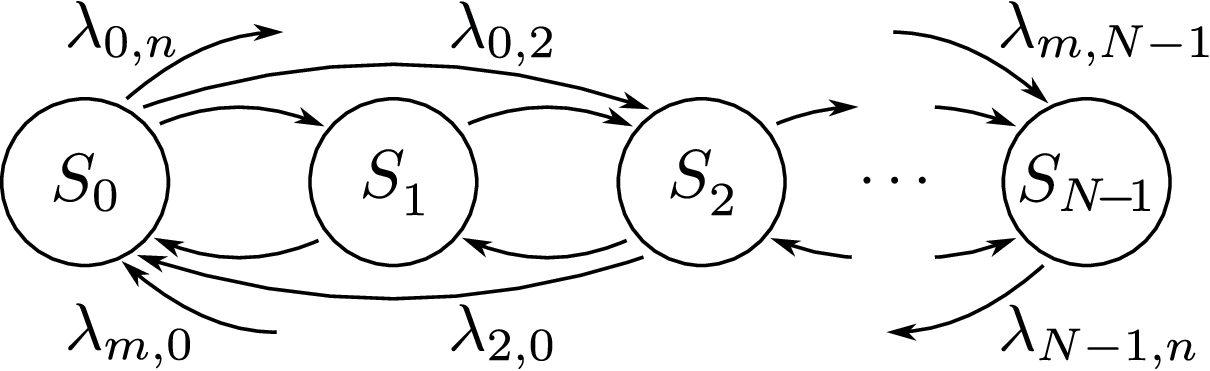}
	\caption{State diagram of the CTMC model with respective transition rates.}
\label{fig:2}
\end{figure}

Since the influence of other vital signs is ignored, it can be assumed that the random variables $\left\lbrace \tau_{\ell} \right\rbrace$ are independent, so that the process $X(t)$ exhibits the memoryless property~\cite{L-G2008}.
Accordingly, the $\ell\text{-th}$ sojourn time, conditioned on state $S_{n}$, has the following exponential distribution:
\begin{equation}
	\tau_{\ell} \sim \exp\left( \mu_{n} \right)
\label{eq:TimesDistribution}
\end{equation}
where $\mu_{n}$ is the parameter of the distribution and can be interpreted as the rate at which the process $X(t)$ leaves the state $S_{n}$~\cite[Sect.~11.4]{L-G2008}.
In Figure~\ref{fig:2}, a generic state diagram of the proposed CTMC is shown, where $\lambda_{m,n}$, $m, n \in \left\lbrace 0,1,\ldots,N-1 \right\rbrace$, denotes the transition rate from state $S_{m}$ to state $S_{n}$.

The statistical behavior of a CTMC is characterized by its infinitesimal generator (or transition rate) matrix~\cite{Papoulis02}

\begin{equation}
	\mathbf{\Lambda} =
	\begin{bmatrix}
		\lambda_{0,0} & \lambda_{0,1} & \cdots & \lambda_{0,N-1}\\
		\lambda_{1,0} & \lambda_{1,1} & \cdots & \lambda_{1,N-1}\\
		\vdots & \vdots & \ddots & \vdots\\
		\lambda_{N-1,0} & \lambda_{N-1,1} & \cdots & \lambda_{N-1,N-1}
	\end{bmatrix}
\label{eq:GenericGenerator}
\end{equation}
where the entries and the parameter in~\eqref{eq:TimesDistribution} are related as
\begin{equation}
	\mu_{m} = -\lambda_{m,m} = \sum_{\substack{n=0\\n \neq m}}^{N-1}\lambda_{m,n}\text{.}
\label{eq:GeneratorProperty1}
\end{equation}
At the end of each waiting time, a state transition occurs: the arrival state is determined according to the transition probabilities of the embedded Markov chain~\cite[Sect.~11.4]{L-G2008}, which can be obtained by the infinitesimal generator matrix $\boldsymbol{\Lambda}$ in~\eqref{eq:GenericGenerator} as follows:
\begin{equation}
	\mathbf{Q} = \mathbf{I} - \left[ \diag \left(\boldsymbol{\Lambda}\right) \right]^{-1} \boldsymbol{\Lambda}
\label{eq:EmbedMatrix}
\end{equation}
where $\mathbf{I}$ is the $N \times N$ identity matrix and $\left[ \diag \left(\boldsymbol{\Lambda}\right) \right]^{-1}$ is the inverse of the diagonal matrix
\begin{equation}
	\diag \left( \boldsymbol{\Lambda} \right) = 
	\begin{bmatrix}
			\lambda_{0,0} & 0 & \cdots & 0\\
			0 & \lambda_{1,1} & \cdots & 0\\
			\vdots & \vdots & \ddots & \vdots\\
			0 & 0 & \cdots & \lambda_{N-1,N-1}
		\end{bmatrix} \text{.}
\end{equation}
As the embedded Markov chain of the CTMC is assumed ergodic, the stationary distribution, described by an $N$-element vector $\boldsymbol{\pi}$, can be obtained solving the following system of linear equations~\cite{L-G2008,Papoulis02}:
\begin{equation}
	\left\lbrace	
	\begin{aligned}
		\boldsymbol{\pi} \, \boldsymbol{\Lambda} & = \boldsymbol{0} \\
		\sum_{n=0}^{N-1}\pi_{n} & = 1
	\end{aligned}
	\right.
\label{eq:LimitingProb}
\end{equation}
where the last equation specifies the normalization of the probability distribution.

Setting appropriately the matrix $\mathbf{\Lambda}$ with values extracted from a real breathing patient, it is possible to completely specify the CTMC model and employ it to replicate a RR pattern with a proper statistical behavior.
Given that each patient can generate different RR patterns, depending on many factors, the infinitesimal generator matrix must be estimated.
For this purpose, the estimation of $\mathbf{\Lambda}$ is carried out in three steps: first, the RR pattern of the patient versus time is estimated; second, the obtained pattern is fitted to a model with $N$ states and, finally, the transition rates are estimated from the finite-state pattern obtained at the previous step.

\subsection{Respiratory Rate Estimation}
\label{subsec:RRestimation}
\begin{figure}[t]
	\centering
	\includegraphics[width=0.75\textwidth]{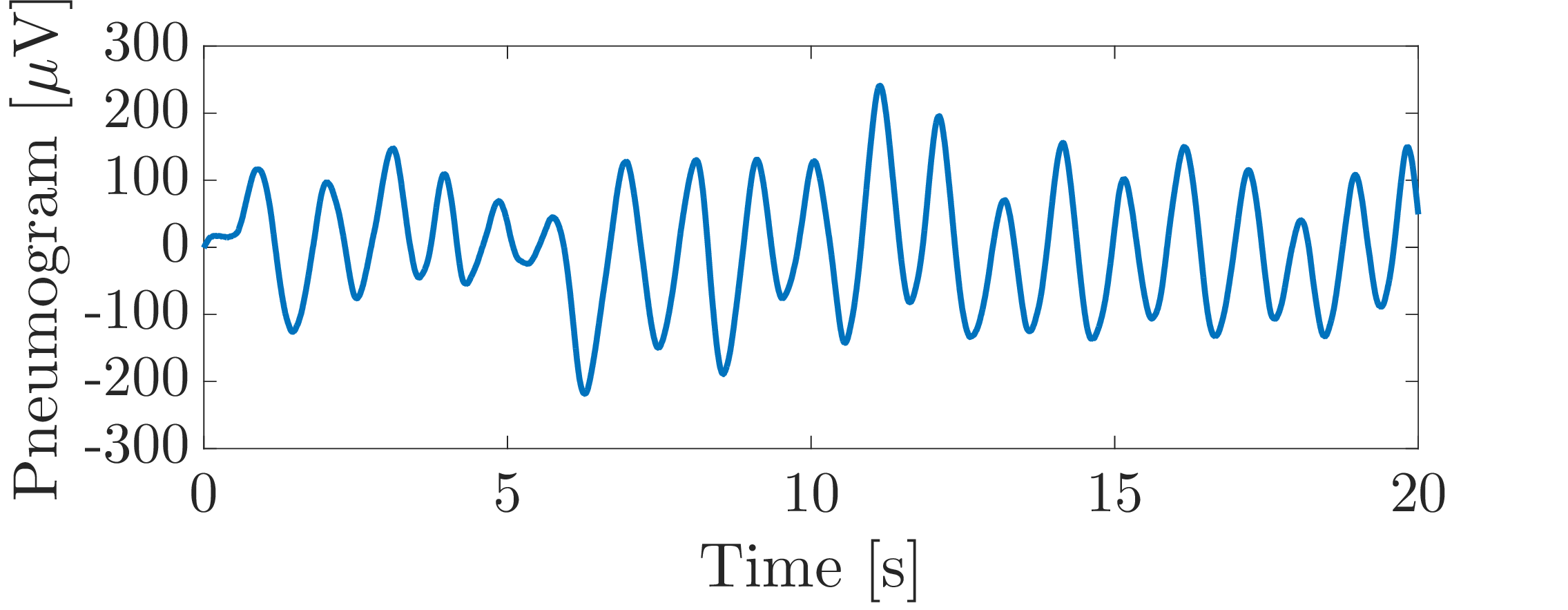}
	\caption{An example of a recorded pneumogram signal of a newborn patient kept under clinical observation.}
\label{fig:2A}
\end{figure}

The RR is estimated by processing the pneumogram signal, which records the volume changes of the thoracic cavity of a patient and is obtained by placing an elastic belt around the chest. In Figure~\ref{fig:2A}, an illustrative example of the pneumogram signal relative to a newborn patient is shown.
Since the pneumogram describes the whole movements related to breathing, an algorithm to estimate the RR from this signal is needed.
Excluding possible respiratory pauses or macro-movements of the patient under observation, the pneumogram signal is a quasi-periodic signal.
A method to estimate the RR from the pneumogram signal thus relies on the estimation of its fundamental frequency.

To this purpose, the pneumogram signal can be modeled by the following discrete-time signal: 
\begin{equation}
	p \left[ i \right] = c+ A\cos\left[ 2 \pi \frac{f_{0}}{f_{s}}i + \phi \right] + w\left[ i \right]
\label{eq:PnemoModel}
\end{equation}
where $c$ is a constant component, $A$ is the amplitude of the periodic component, $f_{0}$ is the fundamental frequency, $f_{s}$ is the sampling frequency, $\phi$ is the phase and $\left\lbrace w[i] \right\rbrace$ is a sequence of independent and identically distributed (i.i.d.) zero-mean Gaussian noise samples.
The main goal is to estimate the fundamental frequency $f_{0}$, which can be interpreted as the RR.

A possible approach to estimate $f_{0}$ relies on the application of the Maximum Likelihood (ML) criterion on a window of signal samples.
Defining the parameter vector $\boldsymbol{\theta} = \left[ A,f_{0},\phi \right]$, the likelihood function $\mathcal{J} \left( \boldsymbol{\theta} \right) $ to be minimized is (see, e.g.,~\cite{Kay93}):
\begin{equation}
	\mathcal{J}\left( \boldsymbol{\theta} \right) = \sum_{i=0}^{M-1}\left( p \left[ i \right] - A\cos\left[ 2 \pi \frac{f_{0}}{f_{s}}i + \phi \right] \right)^{2}
\label{eq:LikeFunction}
\end{equation}
where $\left\lbrace p[i] \right\rbrace_{i=0}^{M-1}$ are now the observed samples in the considered $M$-sample time window.
As a result of algebraic manipulations and proper simplifications~\cite[Sect.~7.10]{Kay93}, an approximate ML fundamental frequency estimator reads:
\begin{equation}
	\widehat{f}_{0} =  \frac{f_{s}}{M} \argmax_{k \in \left\lbrace 0,1,\ldots ,M-1 \right\rbrace} \left| P\left[ k \right] \right|^{2}
\label{eq:FundFreqEstimator}
\end{equation}
where 
\begin{equation}
	P\left[k\right] = \sum_{i=0}^{M-1}p\left[ i \right]e^{-\jmath \frac{2\pi}{M}i}
\label{eq:DiscreteFourierTransform}
\end{equation}
is the discrete Fourier transform of the observed samples.
The set of discrete frequencies $\left\lbrace 0,1,\ldots ,M-1 \right\rbrace$ can be associated with the physical frequencies by the conversion factor $f_{s}/M$.
Once $\widehat{f}_{0}$ has been obtained, it is possible to estimate the amplitude $A$ and determine the magnitude of the periodic component.
Given~\eqref{eq:FundFreqEstimator}, an approximate ML amplitude estimator is~\cite[Sect.~7.10]{Kay93}
\begin{equation}
	\widehat{A} = \frac{2}{M} \left| \sum_{i=0}^{M-1} p \left[ i \right] e^{- \jmath 2\pi \frac{\widehat{f}_{0}}{f_{s}}i} \right| \text{.}
\label{eq:AmpliEstimator}
\end{equation}
If the estimated RR is below the value $R_{L}$ or the amplitude $\widehat{A}$ is lower than a suitable threshold, absence of breathing is assumed and $\widehat{f}_{0}$ is set to $0$.
The frequency $\widehat{f}_{0}$ is finally declared as the RR of the patient in the observed window.

Since the pneumogram signal can be significantly noisy because of possible patient's movements or artifacts involving other body parts, which are associated with state $S_{N-1}$ as described at the beginning of Section~\ref{sec:RRStatModel}, a preliminary time-domain check on the observed samples $\left\lbrace p[i] \right\rbrace_{i=0}^{M-1}$ is performed to detect such conditions.
To this purpose, the analyzed window of samples of the pneumogram signal is checked against the condition
\begin{equation}
\exists i: \; \left| p[i] \right| > \eta \qquad i = 0,1,\ldots ,M-1
\label{eq:PneumoThreshold}
\end{equation}
where $\eta$ is a threshold to distinguish respiratory movements from other ones. Then the estimated frequency $\widehat{f}_{0} = R_{M}$ is formally assigned so that $\varrho_{N-1} = R_{M}$.

\begin{figure}[t]
	\centering
	\includegraphics[width=0.75\textwidth]{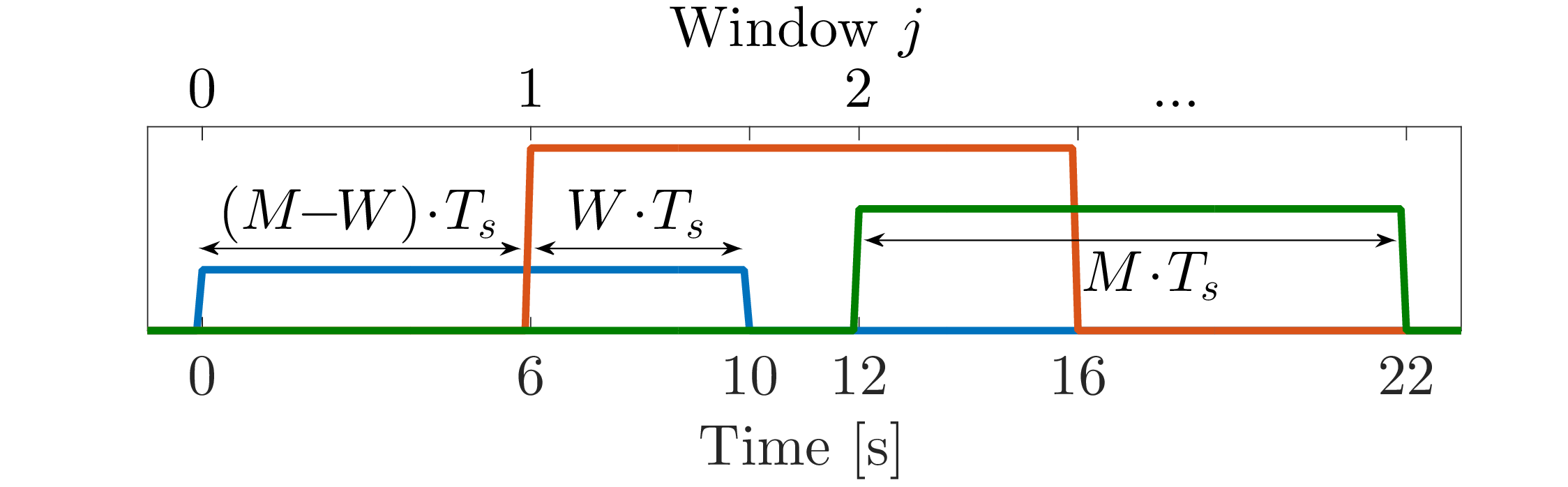}
	\caption{An illustrative example of interlaced windows, with a window length of $M \cdot T_{s} = 10$~s and interlacing interval $W \cdot T_{s} = 4$~s. Window analysis is carried out every $(M-W) \cdot T_{s} = 6$~s.}
\label{fig:2B}
\end{figure}

In order to obtain a RR pattern, which represents the fundamental frequency $\widehat{f}_{0}$ over time, the estimation in~\eqref{eq:FundFreqEstimator} is repeated over successive windows.
Interlaced observation windows, with an interlacing interval of $W$ samples, allow to carry out the estimation every ${M-W}$ samples.
Figure~\ref{fig:2B} shows an example of interlacing, with $ (W/M) \cdot 100 = 40\%$  overlap and three consecutive windows. The integer $j$ specifies the window index.
Estimating the RR along the pneumogram by the approach described above, it is possible to obtain a discrete-time signal $\widehat{X}_{c}\left[ j \right]$, representing the time evolution of the RR, where $\widehat{X}_{c}\left[ j \right]$ is defined as the fundamental frequency estimated in the $j$-th analyzed window by the above procedure. 
Therefore, the RR is estimated every ${\left(M-W\right)\cdot T_{s}}$~seconds, where $T_{s} = 1/f_{s}$ is the sampling interval of the pneumogram.

In order to quantize the estimated continuous-value RR into a finite state space, so that the statistical model introduced in Section~\ref{sec:RRStatModel} can be used, a proper quantization of the values of the continuous-value discrete-time signal $\widehat{X}_{c}\left[ j \right]$ is needed.
This is the focus of the next subsection.

\subsection{Respiratory Rate Signal Quantization}
\label{subsec:RRFitt}
The first step to fit the estimated signal $\widehat{X}_{c}\left[ j \right]$ into the model described in Section~\ref{sec:RRStatModel} is to define the $N$ states in $\mathcal{S}$ by selecting appropriate RR values $\left\lbrace \varrho_{n} \right\rbrace_{n=0}^{N-1}$.
An automatic method to select these values is used, with specific features depending on the presence of apnea events or random large body movements which affect the pneumogram signal.
Following the different modeling cases described at the beginning of Section~\ref{sec:RRStatModel}, different signal quantization methods are considered.
\begin{itemize}
\item If apnea events are not of interest and there are no patient random movements, the $N$ RR values of the model can be selected by the use of the Lloyd-Max algorithm~\cite{MaxQuant60,LLoydQuant82}, which minimizes the mean square distortion between the signal $\widehat{X}_{c}\left[ j \right]$ and the $N$-state quantized one in the range of interest $\left[R_{L},R_{H}\right]$.
\item  If apnea events are of interest, the method is modified as follows: a first state $S_{0}$ with rate ${\varrho_{0} = 0}$ is assigned to describe absence of breathing; the remaining $N-1$ states $\left\lbrace S_{1},S_{2},\ldots,S_{N-1} \right\rbrace$ are estimated by the same Lloyd-Max algorithm used above. In this case, the overall range of interest becomes $\left\lbrace 0 \right\rbrace \cup \left[R_{L},R_{H}\right]$.
\item If random body movements are considered, the algorithm assigns the $N$-th state $S_{N-1}$ with $\varrho_{N-1} = R_{M}$ to time intervals in which the patient is moving; the remaining $N-1$ states $\left\lbrace S_{0},S_{1},\ldots,S_{N-2} \right\rbrace$ are estimated by the Lloyd-Max algorithm. The overall range becomes $\left[R_{L},R_{H}\right] \cup \left\lbrace R_{M} \right\rbrace$.
\item If there are both apneas/respiratory pauses and large random movements, the algorithm assigns the state $S_{0}$ with rate ${\varrho_{0} = 0}$ or the state $S_{N-1}$ with $\varrho_{N-1} = R_{M}$ to time intervals in which absence of breathing is detected or the patient is moving, respectively; the remaining $N-2$ states $\left\lbrace S_{1},S_{2},\ldots,S_{N-2} \right\rbrace$ are estimated by the Lloyd-Max algorithm. The overall range becomes $\left\lbrace 0 \right\rbrace \cup \left[R_{L},R_{H}\right] \cup \left\lbrace R_{M} \right\rbrace$.
\end{itemize}

\begin{figure}[t]
	\centering
	\includegraphics[width=0.75\textwidth]{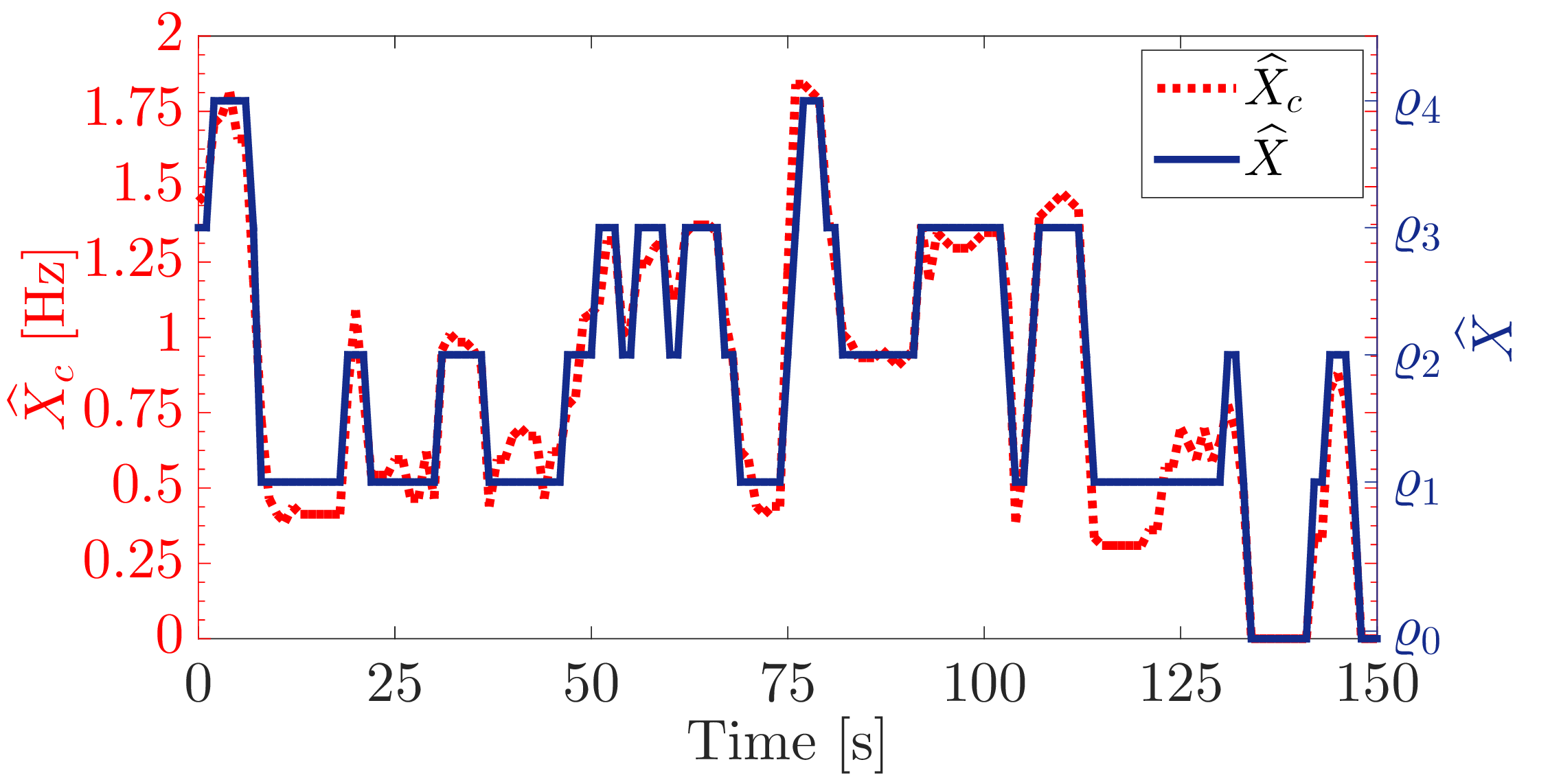}
	\caption{An illustrative example of the RR signal $\widehat{X}_{c}$ estimated from the pneumogram and a quantized model with $N = 5$ states signal $\widehat{X}$; in this examples, a patient with possible respiratory pauses is considered.}
\label{fig:3}
\end{figure}

Once the RRs $\left\lbrace \varrho_{n} \right\rbrace$ and the corresponding states in $\mathcal{S}$ are defined, the pneumogram signal is quantized to the nearest RR value $\varrho_{n}$ present in the model, thus obtaining the following discrete-value version of the signal $\widehat{X}_{c}\left[ j \right]$:
\begin{equation}
	\widehat{X}\left[ j \right] \ \ = \!\! \argmin_{\varrho_{n} \in \left\lbrace \varrho_{0},\varrho_{1},\ldots, \varrho_{N-1} \right\rbrace} \left| \widehat{X}_{c}\left[ j \right] - \varrho_{n} \right| \text{.}
\label{eq:StateApprox}
\end{equation}
In Figure~\ref{fig:3}, an illustrative example of the RR pattern $\widehat{X}_{c}[j]$, estimated by the method described in Section~\ref{subsec:RRestimation}, together with the corresponding quantization $\widehat{X}[j]$, computed according to~\eqref{eq:StateApprox}, are shown.
In this example, the estimation of the RR is carried out every second, on a temporal window of $M \cdot T_{s}=10$~s and with window overlapping equal to $90 \%$ (i.e., $W \cdot T_{s} = 9$~s).

\subsection{Infinitesimal Generator Matrix Estimation}
\label{subsec:InfGenEst}
Given the $N$-state model and the quantized RR pattern extracted from a sample patient, the description of the statistical model of the RR requires the definition of the transition rate matrix $\mathbf{\Lambda}$.
The estimation of the infinitesimal generator matrix of CTMCs is a known problem, which becomes more difficult if the estimation is carried out on incomplete data or on sampled time series~\cite{MDJS2007}. 
To simplify the discussion, $\widehat{X}[j]$, whence the matrix $\mathbf{\Lambda}$ is estimated, is approximated as a continuous-time equivalent signal $\widehat{X}(t)$---this assumption is valid, provided the RR is estimated with sufficiently high frequency, i.e., with a small $\left(M-W\right) \cdot T_{s}$ factor (e.g., $\left(M-W\right) \cdot T_{s} \leq 1$~s).

Owing to this simplification, the estimation of the transition rates is based on the ML estimation method~\cite{MDJS2007,BlaSo2005}.
Assuming the process $\widehat{X}(t)$ is observed in the interval $\left[ 0,T \right]$, the time spent by the process in state $S_{n}$ can be denoted by the random variable $R_{n}\left(T\right)$; similarly, the number of transitions from state $S_{m}$ to state $S_{n}$ in the same observation interval can be denoted as $N_{m,n}\left( T \right)$.
The log-likelihood function to derive an estimate $\widehat{\boldsymbol{\Lambda}}$ of the infinitesimal generator matrix of the observed process $\widehat{X}(t)$ is given by~\cite{BlaSo2005}
\begin{equation}
\log \left[ \mathcal{L}\left( \boldsymbol{\Lambda} \right) \right] =
	\sum_{m=0}^{N-1} \sum_{\substack{n=0\\ n \neq m}}^{N-1} \Big[ N_{m,n}\left(T\right) \log \left(\lambda_{m,n}\right) - \lambda_{m,n}R_{n}\left(T\right) \Big] \text{.}
\label{eq:MLEgenerator}
\end{equation}
By straightforward manipulations~\cite{MDJS2007}, the estimate of $\lambda_{m,n}$ can be expressed as
\begin{equation}
	\widehat{\lambda}_{m,n} = \frac{N_{m,n}\left(T\right)}{R_{n}\left(T\right)} \qquad \text{for } m \neq n\text{.}
\end{equation}
The remaining rates for $m = n$ are obtained using~\eqref{eq:GeneratorProperty1}.

Provided the observation interval $\left[ 0,T \right]$ of the RR pattern is sufficiently long, the ML approach allows a reliable estimation of the infinitesimal generator matrix $\boldsymbol{\Lambda}$.

\section{Simulators}
\label{sec:Simulators}
The CTMC-based statistical model presented in Section~\ref{sec:RRStatModel} can be used to derive appropriate simulators to reproduce respiratory movements and patient-like RR patterns.
These simulators may be useful to test and design video processing algorithms to monitor the RR.
Two simulators are proposed: the \emph{first one} processes a recorded video of a patient breathing normally with an approximately constant RR in order to alter the RR according to the model; the \emph{second one} physically reproduces breathing movements as moving body parts of a manikin.

Simulation of random body movements is typically not of interest and is not considered in this discussion.
If the estimated model includes the state describing patient movements (i.e., the state with $\varrho_{N-1}  = R_{M}$ exists), such a state is excluded and not simulated.
This can be easily achieved by removing the last row and column  of the model infinitesimal generator matrix; then, the diagonal entries are re-calculated according to~\eqref{eq:GeneratorProperty1}.

As a first step, both simulators generate sojourn times for the finite state model.
Assuming that the initial state is unknown, it can be randomly drawn according to the stationary distribution of the CTMC defined in~\eqref{eq:LimitingProb}.
Then, the first sojourn time $\tau_{1}$ in the initial state $S_{m}$ is generated according to the proper exponential distribution~\eqref{eq:TimesDistribution} with parameter $\mu_{m}$.
At the end of the first sojourn time, a state transition occurs: the arrival state is randomly drawn according to the distribution extracted from the $m\text{-th}$ row of the matrix $\mathbf{Q}$ in~\eqref{eq:EmbedMatrix}.
The simulation of sojourn times $\left\lbrace \tau_{\ell} \right\rbrace$ and state transitions continues until the desired duration of the simulation is reached.
The obtained sojourn times $\left\lbrace \tau_{\ell} \right\rbrace$ are used in the simulators to generate the RR pattern.

\subsection{Video-Based Simulator}
\label{subsec:VidSim}
The video-based simulator processes a video of a regularly breathing patient and modifies it by creating a new video with variable RRs.
The RR of the framed patient in the source video is approximated as time-invariant and defined as $\varrho_{V}$.
The rates associated with the various states can be described by the ratios between the rates of the corresponding state and the RR of the patient in the original video. 
The normalized RRs in the model can be expressed as:
\begin{equation}
	\bar{\varrho}_{n} = \frac{\varrho_{n}}{\varrho_{V}} \qquad n = 0,1,\ldots,N-1 \text{.}
\label{eq:NormRates}
\end{equation}
After the waiting and jump times are generated according to the $N$-state CTMC, the corresponding number of frames are obtained as follows:
\begin{align}
	\tilde{\tau}_{\ell} &= \mathrm{round}\left( \tau_{\ell} \cdot f_{r} \right) \\
	\tilde{t}_{\ell} &= \mathrm{round}\left( t_{\ell} \cdot f_{r} \right)
\end{align}
where $f_{r}$ is the frame rate of the video input and $\mathrm{round}\left(\cdot\right)$ denotes the integer closest to the argument (rounding function).
If the RRs $\left\lbrace \varrho_{n} \right\rbrace$ are all different from the rate $\varrho_{V}$ of the original video, it may be convenient to scale them by a factor $C_{V}$ chosen so that $\varrho_{\mathrm{max}} \cdot C_{V} = \varrho_{V}$, where $\varrho_{\mathrm{max}}$ is the RR corresponding to the state with lowest rate (${\min_{n \in \{0,1,\ldots,N-1\}} \mu_{n}}$) and, consequently, longest mean sojourn time.

The simulator then starts producing a new video where the artificially generated RR pattern is inserted.
The system scans the whole video inserting breathing times with durations $\left\lbrace \tilde{\tau}_{\ell} \right\rbrace$.
The $\ell\text{-th}$ breathing time is simulated by processing a video block of $\tilde{d}_{\ell}$ frames starting from the $\tilde{t}_{\ell}\text{-th}$ video frame, where
\begin{equation}
	\tilde{d}_{\ell} = \left\lceil \frac{\tau_{\ell}}{\bar{\varrho}_{n}} \cdot f_{r} \right\rceil
\end{equation}
and $\left\lceil \cdot \right\rceil$ denotes the smallest integer larger than the argument (ceiling function).
 A RR is generated according to the state to be simulated:
\begin{itemize}
	\item if $S_{n}$ is such that $\bar{\varrho}_{n} = 1$, no video processing is needed, since the desired RR is equal to that of the patient breathing in the original video;
	\item if $S_{n}$ is such that $0 < \bar{\varrho}_{n} < 1$, a block of $\tilde{d_{\ell}} < \tilde{\tau}_{\ell}$ frames is extracted, in order to slow down breathing movements of the recorded patient; the extracted video block is ``stretched'' to the proper length $\tilde{\tau}_{\ell}$ by the use of pixel-wise interpolation in the temporal dimension by a cubic spline~\cite{CubSpline2000};
	\item if $S_{n}$ is such that $\bar{\varrho}_{n} > 1$, a block of $\tilde{d_{\ell}} > \tilde{\tau}_{\ell}$ frames is extracted, in order to speed up breathing movements of the recorded patient, ``contracting'' the video block to the proper length $\tilde{\tau}_{\ell}$ by decimation in the temporal dimension using a cubic spline~\cite{CubSpline2000};
	\item if $S_{0}$ is such that $\bar{\varrho}_{0} = 0$, the procedure used for $0 < \bar{\varrho}_{n} < 1$ cannot be used. In this case, the value of $\bar{\varrho}_{0}$ is replaced by a new value $\bar{\varrho}'_{0}$ properly chosen such that $0 < \varrho'_{0} \ll R_{L}$, where $\varrho'_{0} = \bar{\varrho}'_{0} \cdot \varrho_{V}$, so that the simulated RR is a convenient value clinically considered an apnea.
	Then, the procedure illustrated for $0 < \bar{\varrho}_{n} < 1$ is applied with a modified $\bar{\varrho}'_{0}$ value.
	The results described in the next section use a value $\varrho'_{0} \simeq 0.1$~Hz.
\end{itemize}

For $\bar{\varrho}_{n} < 1$, the noise, present in the original video and caused by camera sensors, wired connections and the environment, is also subject to interpolation.
This means that the noise ``slows down'' or, more precisely, is filtered by the interpolating filter.
In order to maintain noise characteristics similar to those of the original video, a noise compensation algorithm has been devised.
Assuming that the noise is uniformly distributed in each frame, its statistical characterization is time-invariant and can be modeled by additive white Gaussian noise, whose characteristics are estimated by processing pixels with static background.
Selecting a background pixel region and assuming a static video-camera, the system first determines an estimate  $\hat{\sigma}^{2}$ of the noise variance by averaging the estimated variances for every pixel in the considered region.
In this process, each pixel variance is estimated by the sample variance~\cite{Papoulis02}.
To compensate for the noise not included because of the interpolation filter, a sequence of uncorrelated Gaussian samples with zero mean and variance $\hat{\sigma}^{2}$ is generated and then filtered by a high-pass filter with $3$~dB cut-off frequency equal to that of the interpolating filter.
To update the camera noise inside the new video block, the filtered noise sequence is added to all time-interpolated pixels.
This procedure is not necessary for the cases with $\bar{\varrho}_{n} \geq 1$, because the decimation process does not modify the noise statistic.
The overall procedure is repeated until the complete RR pattern has been inserted into the video.

Video simulation examples of breathing newborns with varying RRs and possible apnea episodes are provided as supplemental materials following the descriptions in Subsections~\ref{subsec:ModelValidation} and~\ref{subsec:ApneaSim}.
\begin{figure}[t!]
	\centering
	\includegraphics[width=0.6\textwidth]{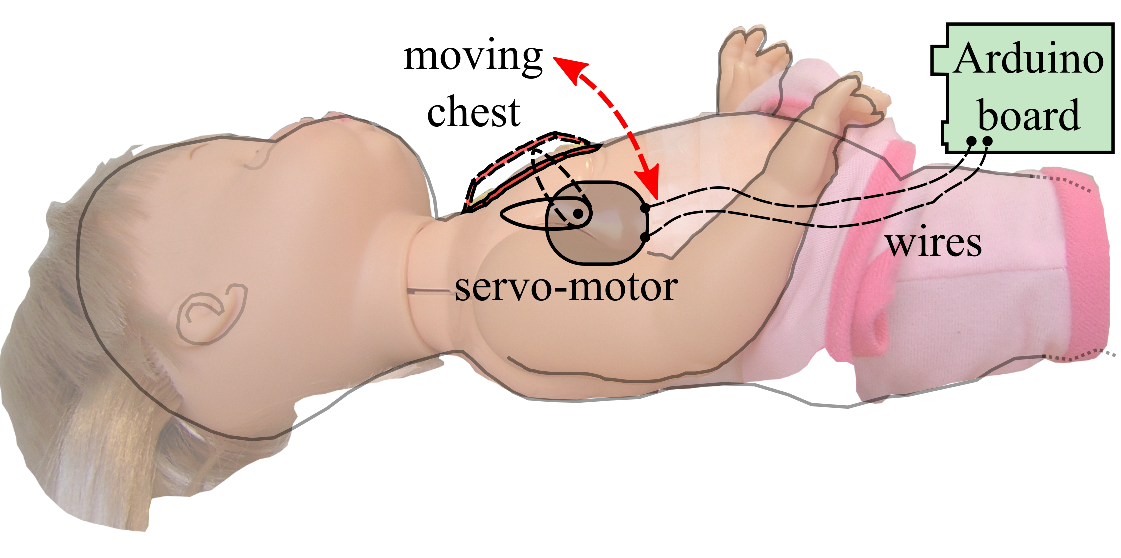}
	\caption{Illustrative diagram of the hardware-based simulator. The moving chest is driven by the servo-motor inside the manikin connected to a motor shield for an Arduino board.}
\label{fig:4}
\end{figure}

\subsection{Hardware-Based Simulator}
\label{subsec:HardSim}
The hardware-based simulator consists of a manikin of an infant with moving parts able to reproduce respiratory chest movements of the newborn.
In Figure~\ref{fig:4}, an illustrative representation of the manikin simulator is shown.
It consists of a moving chest coupled with the body of the manikin.
The moving chest is driven by a mechanical arm connected to an electric servo-motor, inserted inside the body of the manikin, and controlled by a motor shield for Arduino UNO~\cite{Arduino13}, a board based on an Atmel ATmega328P microcontroller.
The controller is able to move the chest at a user-defined frequency with asymmetric speed to distinguish inhalation and exhalation movements.
The servo-motor can vary the RR approximately from $2$ bpm to $200$ bpm, equivalent to a RR range between $0.033$ Hz and $3.33$ Hz, which readily allows to simulate the RRs of a newborn.

To simulate the respiratory behavior of a newborn, times $\left\lbrace \tau_{\ell} \right\rbrace$ and corresponding states with RRs $\left\lbrace \varrho_{n} \right\rbrace$ generated by CTMC-based simulation are passed to the microcontroller, which drives the servo-motor to mimic the RR pattern, moving the chest of the manikin with the selected RR for the required time.
In the case of a respiratory pause or apnea event, the servo-motor is slowed down to the minimum reachable rate for a time duration equal to the sojourn time in the apnea state.

\section{Applications and Results}
\label{sec:ApplRes}
\subsection{Model Validation and Simulators}
\label{subsec:ModelValidation}
First, the validation of the CTMC model is discussed. Vital sign recordings were provided by the Neonatal Intensive Care Unit (NICU) of the University Hospital of Parma.
As examples, two recorded pneumogram samples of two different newborns, the first suffering from apnea events and the second regularly breathing, are used for the RR data extraction: the first record has a total length of $1$~h and $42$~min and the second one of $1$~h and $6$~min.
The fundamental frequency is estimated on temporal windows of duration $M\cdot T_{s} = 10$~s (with $T_{s} = 31.25$~ms) and with a $95\%$ overlapping, namely $W\cdot T_{s} = 9.5$~s.
Moreover, a heuristic threshold $\eta = 400$~$\mathrm{\mu V}$ is set in~\eqref{eq:PneumoThreshold} to detect random movements to be assigned to the state $S_{N-1}$---the selected value of $\eta$ has been experimentally optimized. This value is selected by observation of pneumogram signals of several newborn patients.
Fixing, as an example, the desired number of states of the model to $N = 5$, the automatic state selector, described in Subsection~\ref{subsec:RRFitt}, extracts the RR sets $\left\lbrace 0, 0.5, 0.9, 1.3, R_{M} \right\rbrace$ and $\left\lbrace 0.44, 0.74, 1.04, 1.33, R_{M} \right\rbrace$ for the first and the second considered patients, respectively.
Afterward, relying on the methods described in Section~\ref{sec:RRStatModel}, the algorithm derives an estimate $\widehat{\boldsymbol{\Lambda}}$ of the infinitesimal generator matrix.
In Tables~\ref{tab:1A} and~\ref{tab:1B} the RR sets $\left\lbrace \varrho_{n} \right\rbrace$, the ML-estimated matrices $\widehat{\boldsymbol{\Lambda}}$ and the corresponding stationary distributions $\boldsymbol{\pi}_{\mathrm{ML}}$ derived according to~\eqref{eq:LimitingProb} for the two considered examples, are reported.
It can be noticed that both patients are affected by random body movements which cause the presence of the state with $\varrho_{N-1} = R_{M}$.
\begin{table*}[t!]
	\renewcommand{\arraystretch}{1.1}
	\centering
	\begin{subtable}[t]{\linewidth}
		\centering
		\footnotesize
		\begin{tabular}{c}
			\hline
			\textbf{Example 1 (patient suffering from apnea events)}\\
			\hline
			$\left\lbrace \varrho_{n} \right\rbrace = \begin{Bmatrix}
									0 & 0.5 & 0.9 & 1.32 & R_{M}
									\end{Bmatrix}$\\
			\hline
	%		\noalign{\vskip 0.5mm}
			$\widehat{\boldsymbol{\Lambda}} = \begin{bmatrix}
						-0.188785 & 0.08972 & 0.04486 & 0.042991 & 0.011215\\
						0.047083 & -0.203685 & 0.073695 & 0.071648 & 0.011259\\
						0.014614 & 0.080376 & -0.22547 & 0.127349 & 0.003132\\
						0.007717 & 0.019756 & 0.036734 & -0.066677 & 0.00247\\
						0.042272 & 0.02642 & 0.002642 & 0.002642 & -0.073976
					\end{bmatrix}$\\
	%		\noalign{\vskip 0.5mm}
			\hline
			$\boldsymbol{\pi}_{\mathrm{ML}} = \begin{bmatrix}
												0.08788 & 0.16048 & 0.15736 & 0.53211 & 0.06217
											\end{bmatrix}$\\
			\hline
		\end{tabular}
		\caption{}
		\label{tab:1A}
	\end{subtable}
	\vskip 0.5mm
	\begin{subtable}[t]{\linewidth}
		\centering
		\footnotesize
		\begin{tabular}{c}
			\hline
			\textbf{Example 2 (patient regularly breathing)} \\
			\hline
			$\left\lbrace \varrho_{n} \right\rbrace = \begin{Bmatrix}
											0.44 & 0.74 & 1.04 & 1.33 & R_{M}
									\end{Bmatrix}$\\
			\hline
		%	\noalign{\vskip 0.5mm}
			$\widehat{\boldsymbol{\Lambda}} = \begin{bmatrix}
								-0.205567 & 0.124197 & 0.059957 & 0.004283 & 0.017131\\
								0.020036 & -0.080859 & 0.047943 & 0.003578 & 0.009302\\
								0.014957 & 0.071581 & -0.108974 & 0.014957 & 0.007479\\
								0.015873 & 0.047619 & 0.206349 & -0.317460 & 0.047619\\
								0.003656 & 0.007313 & 0.006399 & 0 & -0.017367		
							\end{bmatrix}$\\
			%\noalign{\vskip 0.5mm}
			\hline
			$\boldsymbol{\pi}_{\mathrm{ML}} = \begin{bmatrix}
												0.05644 & 0.32998 & 0.22677 & 0.01516 & 0.37164
											\end{bmatrix}$\\
			\hline
	\end{tabular}
	\caption{}
	\label{tab:1B}
	\end{subtable}
	\caption{Estimated RRs sets, infinitesimal generator matrices and corresponding stationary distributions of (a) a newborn patient suffering from apnea events and (b) a newborn regularly breathing.}
	\label{tab:1}
\end{table*}

It must be remarked that the description of large body movements or possible artifacts with a specific state with rate $\varrho_{N-1} = R_{M}$ in the CTMC is fundamental in order to avoid degradation of the statistical behavior of the model.
In fact, if this state was not available, faulty estimation of the RRs might arise, with possible wrong selection of RR values $\left\lbrace \varrho_{n} \right\rbrace$ and incorrect estimates of transition rates and infinitesimal generator matrix.

\begin{figure}[t!]
	\centering
	\includegraphics[width=0.999\textwidth]{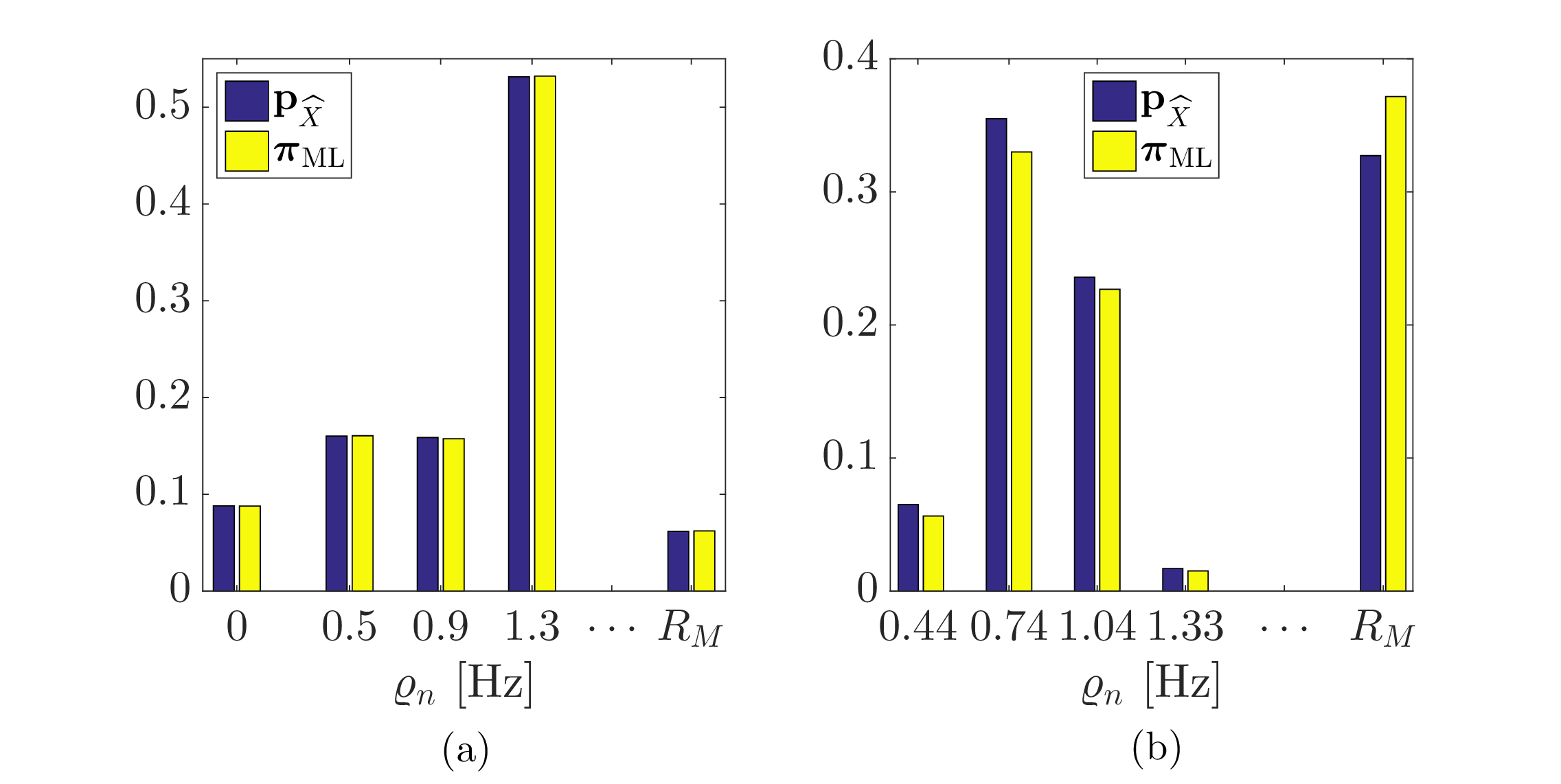}
	\caption{Comparisons between histograms with automatically selected bins for the process $\widehat{X}$ (pneumogram-based) with PMF $\mathbf{p}_{\widehat{X}}$ and the stationary distributions  $\boldsymbol{\pi}_{\mathrm{ML}}$ relative to (a) the example of the newborn patient suffering from apnea events and (b) the regularly breathing newborn.}
\label{fig:5}
\end{figure}

In Figure~\ref{fig:5}, comparisons between the histograms of estimated frequencies~$\widehat{X}\left[ j \right]$, whose probability mass function (PMF) is defined as the vector $\mathbf{p}_{\widehat{X}}$, and the stationary distributions $\boldsymbol{\pi}_{\mathrm{ML}}$ relative to the ML-estimated matrix~$\widehat{\boldsymbol{\Lambda}}$ for both examples are shown: Part~(a) corresponds to the first example of the patient suffering from apneas and Part~(b) is related to the second example of the regularly breathing newborn.
In order to quantitatively compare the similarity of the two PMFs, the Kullback-Leibler (KL) divergence~\cite{EIT} may be used:
\begin{equation}
D_{\mathrm{KL}} \left( \mathbf{p} \| \mathbf{q} \right) = \sum_{n}p_{n} \log_{2}\frac{p_{n}}{q_{n}}
\end{equation}
in which $p_{n}$ and $q_{n}$ denote the probability masses of the distributions $\mathbf{p}$ and $\mathbf{q}$, respectively.
This quantity is a measure of the difference between the ``true'' distribution $\mathbf{p}$ and the ``assumed'' distribution $\mathbf{q}$; it is expressed in bits, due to the use of $\log_{2}$.
In both examples shown in Figure~\ref{fig:5}, the KL divergence $D_{\mathrm{KL}} \left( \mathbf{p}_{\widehat{X}} \| \boldsymbol{\pi}_{\mathrm{ML}} \right)$ is computed: values of $0.011 \cdot 10^{-3}$~bits and of $6.642 \cdot 10^{-3}$~bits are obtained for the cases~(a) and (b), respectively.
The stationary distributions are very similar to the histogram-based ones, with very low KL divergence values in both examples, confirming that the CTMC model has a steady state behavior similar to that of the RR pattern of the real patients.

To further verify the effectiveness of the statistical model, the effects of varying the number of states of the CTMC is now discussed.
For this validation example, a new patient is considered: the total length of the recorded pneumogram signal is $51$~min.
In Tables~\ref{tab:2A}, \ref{tab:2B} and~\ref{tab:2C} the extracted RR sets $\left\lbrace \varrho_{n} \right\rbrace$, corresponding estimated infinitesimal generator matrices and stationary distributions $\boldsymbol{\pi}_{\mathrm{ML}} $ with $N = 4$, $5$ and~$6$ are reported, respectively.
Changes in the transition rates when the number of states of the model is varied can be appreciated---the use of a larger number of levels allows a finer representation of the RR values.
\begin{table*}[t]
	\renewcommand{\arraystretch}{1.1}
	\centering
	\begin{subtable}[t]{\linewidth}
		\centering
		\footnotesize
		\begin{tabular}{c}
			\hline
			\textbf{CTMC model with $N = 4$}\\
			\hline
			$\left\lbrace \varrho_{n} \right\rbrace = \begin{Bmatrix}
							0.49 & 0.88 & 1.27 & R_{M}
					\end{Bmatrix} $\\
			\hline
			$\widehat{\boldsymbol{\Lambda}} = \begin{bmatrix}
					-0.196532 & 0.109827 & 0.023121 & 0.063584\\
					0.021876 & -0.058898 & 0.012621 & 0.024401\\
					0 & 0.105263 & -0.144044 & 0.038781\\
					0.006814 & 0.021124 & 0.003407 & -0.031346
					\end{bmatrix}$\\
			\hline
			$\boldsymbol{\pi}_{\mathrm{ML}} = \begin{bmatrix}
															0.0605 & 0.38942 & 0.05555 & 0.49453
														\end{bmatrix}$\\
			\hline
		\end{tabular}
		\caption{}
		\label{tab:2A}
	\end{subtable}
	\vskip 0.5mm
	\begin{subtable}[t]{\linewidth}
		\centering
		\footnotesize
		\begin{tabular}{c}
			\hline
			\textbf{CTMC model with $N = 5$}\\
			\hline
			$\left\lbrace \varrho_{n} \right\rbrace = \begin{Bmatrix}
								0.44 & 0.73 & 1.03 & 1.32 & R_{M}
							\end{Bmatrix}$\\
			\hline
%			\noalign{\vskip 0.5mm}
			$\widehat{\boldsymbol{\Lambda}} = \begin{bmatrix}
					-0.227545 & 0.083832 & 0.023952 & 0.023952 & 0.095808\\
					0.011161 & -0.071429 & 0.035714 & 0.001116 & 0.023438\\
					0.010604 & 0.067869 & -0.123012 & 0.012725 & 0.031813\\
					0 & 0.021978 & 0.076923 & -0.131868 & 0.032967\\
					0.003406 & 0.014986 & 0.011580 & 0.001362 & -0.031335
					\end{bmatrix}$\\
%			\noalign{\vskip 0.5mm}
			\hline
			$\boldsymbol{\pi}_{\mathrm{ML}} = \begin{bmatrix}
															0.02902 & 0.29355 & 0.15485 & 0.02781 & 0.49477
														\end{bmatrix}$\\
			\hline
		\end{tabular}
		\caption{}
		\label{tab:2B}
	\end{subtable}
	\vskip 0.5mm
	\begin{subtable}[t]{\linewidth}
		\centering
		\footnotesize
		\begin{tabular}{c}
			\hline
			\textbf{CTMC model with $N = 6$} \\
			\hline
			$\left\lbrace \varrho_{n} \right\rbrace = \begin{Bmatrix}
								0.41 & 0.65 & 0.88 & 1.11 & 1.35 & R_{M}
							\end{Bmatrix}$\\
			\hline
			\noalign{\vskip 0.25mm}
			$\widehat{\boldsymbol{\Lambda}} = \begin{bmatrix}
					-0.263566 & 0.093023 & 0.015504 & 0.015504 & 0.031008 & 0.108527\\
					0.015102 & -0.133765 & 0.084142 & 0.006472 & 0.002157 & 0.025890\\
					0.005735 & 0.061649 & -0.131900 & 0.035842 & 0 & 0.028674\\
					0.007828 & 0.003914 & 0.101761 & -0.164384 & 0.027397 & 0.023483\\
					0 & 0 & 0.067227 & 0.100840 & -0.201681 & 0.033613\\
					0.002723 & 0.008850 & 0.013615 & 0.005446 & 0.000681 & -0.031314
					\end{bmatrix}$\\
			\noalign{\vskip 0.25mm}
			\hline
			$\boldsymbol{\pi}_{\mathrm{ML}} = \begin{bmatrix}
															0.02143 & 0.1547 & 0.22708 & 0.08513 & 0.01818 & 0.49348
														\end{bmatrix}$\\
			\hline
		\end{tabular}
		\caption{}
		\label{tab:2C}
	\end{subtable}
	\caption{Estimated RR sets and infinitesimal generator matrices for a newborn patient with different values of the number of states $N$.}
	\label{tab:2}
\end{table*}

The simulators described in Section~\ref{sec:Simulators} have been used to generate videos of breathing patients with a statistically-defined RR pattern.
The video-based simulator is used to process video streams recorded in the NICU of the University Hospital of Parma, with cameras having a frame rate $f_{r} = 25$~frame/s. Videos of the hardware-based simulator (i.e., the ``breathing'' manikin) are instead recorded with cameras operating at ${f_{r} = 15}$~frame/s.
Both simulators are employed with the state set $\mathcal{S}$ with RRs set $\left\lbrace \varrho_{n} \right\rbrace$ and the matrix $\widehat{\boldsymbol{\Lambda}}$ extracted as described in the previous section.

These simulators have been used to test and evaluate the performance of video processing-based systems to monitor the RR and detect apnea events.
The obtained video sequences have been analyzed using algorithms developed in previous works.
In particular, the algorithm described in~\cite{BIOMS14} uses a Motion Magnification technique to enhance small breathing movements for Apnea Detection and is here referred to as MMAD.
The algorithm described in~\cite{MEMEA15}, here referred to as Spatio-Temporal video-processing for RR Estimation (STRE), includes a spatio-temporal video processing system to reinforce RR estimation and apnea detection.
Both MMAD and STRE algorithms can extract signals representative of breathing motion from a video stream---they are then used to detect apnea events~\cite{BIOMS14} or to estimate the RR~\cite{MEMEA15}.
These algorithms analyze extracted breathing signals on temporal windows of $10$~s with window interlacing equal to $90\%$. A detailed description of these algorithms is out of the scope of this paper: the interested reader is referred to~\cite{BIOMS14} and~\cite{MEMEA15}.

\subsection{Analysis of RR Estimators by Simulated Breathing}
\label{subsec:SimAnalysisRR}
Video processing-based RR estimators can be tested comparing estimated rates with the ones simulated by the statistical model.
The performance of the STRE algorithm in RR estimation is here assessed considering two performance metrics: (i) the Root Mean Square Error (RMSE) between the simulated rate and that estimated by the video processing-based algorithm and (ii) the probability of correct estimation of RR, defined, according to medical practice, by the condition that the RR falls inside a tolerance range of $\pm 15\%$ with respect to the correct true value.
First, a video with a total length of $17$~min and $54$~s, generated by the software-based simulator, is analyzed.
The simulated video is obtained by processing a video sample of a sleeping newborn breathing regularly with an approximate rate $\varrho_{V} = 0.69$~Hz.
The RR is correctly estimated in $940$ out of $1074$ temporal windows (i.e., with a probability of correct estimation equal to $0.875$), with a RMSE equal to $0.063$~Hz.
By normalizing the RMSE with respect to the average value of the simulated RR, one finds an average relative error of~$10.2\%$.
Then, a video sample of the hardware-based simulator is analyzed: the recording has a total length of $8$~min and $48$~s, during which the manikin simulates a breathing newborn.
In this case, the RR is correctly estimated in $460$ out of $528$ temporal windows (i.e., with a probability of correct estimation equal to $0.871$), with a RMSE equal to $0.083$~Hz, which, normalized with respect to the average value of simulated RR, corresponds to an average relative error of~$9.7\%$.

\subsection{Simulation of Apnea Episodes}
\label{subsec:ApneaSim}
Clinically, an apnea event is defined as an episode of absence of breathing lasting at least $20$~s or between $10$~s and $20$~s, if it is associated with other clinical signs or symptoms~\cite{WYKA2011}.
In the following, adopting a conservative approach, these two conditions are not distinguished and any episode of absence of breathing of at least $10$~s is considered as apnea.
The absence of breathing for less than $10$~s is considered as a respiratory pause and is not clinically relevant.
The statistical model described in Section~\ref{sec:RRStatModel} can be used to simulate apnea episodes and respiratory pauses, in both software- and hardware-based simulators, provided that the state $S_{0}$ is associated with the rate $\varrho_{0} = 0$.
A simple statistical model is a two-state CTMC ($N = 2$), where states $\{S_{0},S_{1}\}$ describe the presence of an apnea/respiratory pause and regular breathing, respectively.
Corresponding sojourn times $\left\lbrace \tau_{\ell} \right\rbrace$ denote the durations of respiratory pauses and normal breathing, conditionally on the corresponding state.
The software-based simulator operates on a video of a regularly breathing patient to produce a new video with artificially introduced respiratory pauses.
The estimation method of the $2 \times 2$ infinitesimal generator matrix and the video-based simulator described in Subsection~\ref{subsec:VidSim} are used with only two states: $\bar{\varrho}_{n} = 1$ if regular breathing occurs and $\bar{\varrho}_{n} = 0$ in the presence of apneas/respiratory pauses.
\begin{figure}[t]
	\centering
	\includegraphics[width=0.75\textwidth]{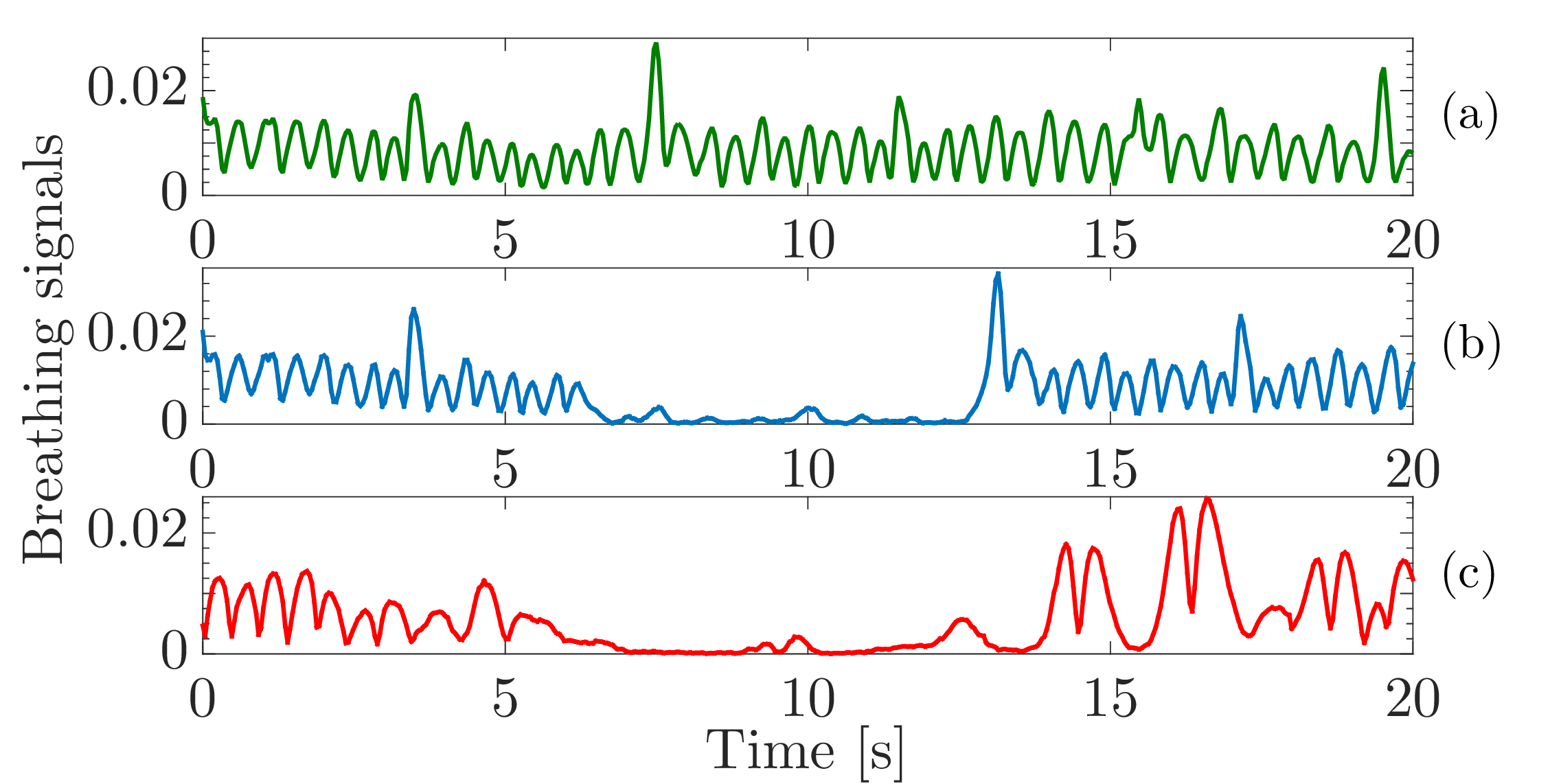}
	\caption{Examples of breathing signals of a newborn: (a)~normal breathing pattern, (b)~a software-simulated respiratory pause and (c) a real respiratory pause in the same patient, lasting approximately as the simulated one.}
\label{fig:6}
\end{figure}

In Figure~\ref{fig:6}, a direct comparison between motion signals extracted from the three possible videos using the STRE algorithm are depicted.
In Part~(a), the breathing signal extracted from a newborn without respiratory pauses is shown, where a RR approximately equal to $1.17$~Hz can be recognized.
In Part~(b), the breathing signal, obtained by inserting a simulated respiratory pause into the video stream related to Part~(a), is shown. In this example, the simulated pause begins at time instant $7.3$~s and lasts approximately $6.2$~s.
For comparison purposes and to demonstrate the effectiveness of the software-based simulator, in Part~(c) the breathing signal extracted from the video of a child suffering from CCHS, containing a real respiratory pause lasting approximately $6.26$~s, is shown.
The similarity of the breathing signal in Figure~\ref{fig:6}(c) with the signal embedding the simulated respiratory pause in Figure~\ref{fig:6}(b) can be appreciated.

\subsection{Analysis of Apnea Detectors by Simulated Breathing}
\label{subsec:SimAnalysisApnea}
Finally, the two simulators have been used to generate videos of newborns suffering from apnea events to test the previously proposed algorithms~\cite{BIOMS14},~\cite{MEMEA15}.
The software-based simulator has been used to generate a video lasting approximately $1$~h.
The obtained video includes $74$ simulated respiratory pauses, with $13$ events, lasting at least $10$~s each, which can be interpreted as apneas.
The total duration of simulated apnea events is $166$~s, with an average duration of $14.55$~s and a maximum duration of $35$~s.
The hardware-based simulator is used to record a video lasting approximately $46$~min.
The simulation includes $33$ simulated respiratory pauses, with $12$ events lasting at least $10$~s each.
The total duration of simulated apnea events is $220$~s, with an average duration of $17.08$~s and a maximum duration of $33$~s.

The obtained videos are processed by the algorithms MMAD~\cite{BIOMS14} and STRE~\cite{MEMEA15}, which implement automatic apnea detection systems, introduced at the end of Subsection~\ref{subsec:ModelValidation}.
As described at the beginning of Subsection~\ref{subsec:ApneaSim}, the two algorithms focus only on events lasting at least $10$~s.
The performance of these detection systems is investigated considering a binary test, which classifies results as ``presence of apnea'' (positive event) or ``normal breathing'' (negative event).
The performance results are presented in terms of sensitivity and specificity~\cite{LM08}, defined, respectively, as
\begin{align}
	\alpha &\triangleq \frac{T_{\mathrm{TP}}}{T_{\mathrm{TP}} + T_{\mathrm{FN}}}\\
	\beta &\triangleq \frac{T_{\mathrm{TN}}}{T_{\mathrm{TN}} + T_{\mathrm{FP}}}
\label{eq:SESP}
\end{align}
where $T_{\mathrm{TP}}$, $T_{\mathrm{TN}}$, $T_{\mathrm{FP}}$, and $T_{\mathrm{FN}}$ denote, respectively, the total length of the time intervals with apnea correctly detected (True Positives), regular breathing correctly detected (True Negatives), regular breathing incorrectly reported as apnea (False Positives) and apnea incorrectly reported as normal breathing (False Negatives).
Sensitivity and specificity can be interpreted, respectively, as the fraction of apnea time which is correctly classified and the fraction of regular breathing time which is correctly identified.
As a global measure of test performance, the Diagnostic Odds Ratio (DOR)~\cite{DOR2003} can also be employed, defined as
\begin{equation}
	\Delta \triangleq \frac{T_{\mathrm{TP}}/T_{\mathrm{FN}}}{T_{\mathrm{FP}}/T_{\mathrm{TN}}} = \frac{\alpha}{1-\alpha} \cdot \frac{\beta}{1-\beta} \text{.}
\label{eq_DOR}
\end{equation}

\begin{figure}[t!]
	\centering
	\includegraphics[width=0.75\textwidth]{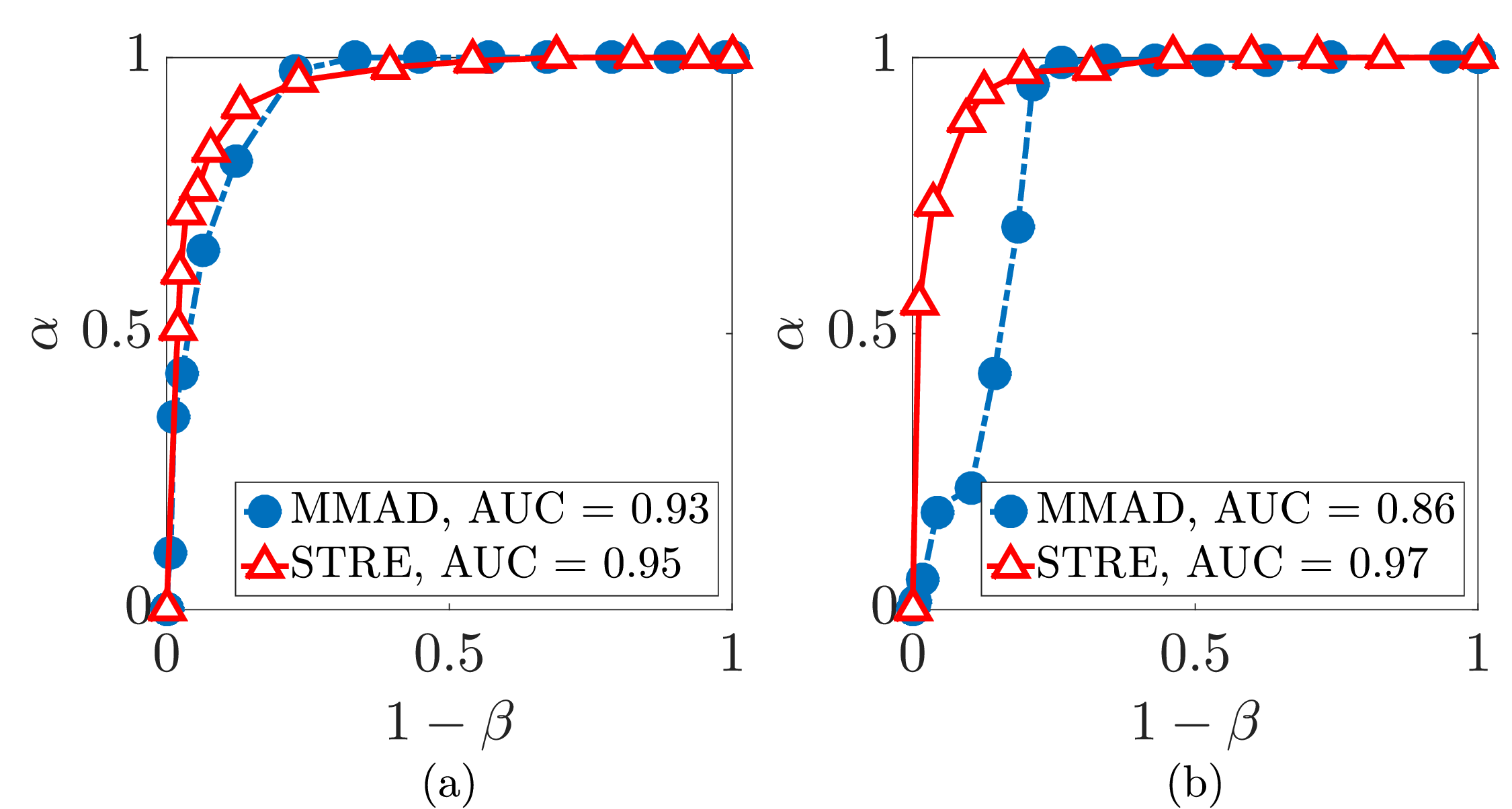}
	\caption{ROC curves for algorithms MMAD and STRE: (a) video-based simulator and (b) hardware-based simulator.}
\label{fig:7}
\end{figure}
In Figure~\ref{fig:7}, the performance results, in terms of Receiver Operating Characteristic (ROC) curves~\cite{ROC88}, are presented.
The curves are obtained testing the algorithms for various values of the decision thresholds for presence/absence of periodicity related to the breathing signal~\cite{BIOMS14,MEMEA15}.
Specifically, in Figure~\ref{fig:7}(a) and Figure~\ref{fig:7}(b) the ROC curves for MMAD and STRE algorithms tested on videos generated by software-based and hardware-based simulators are shown, respectively.
As concise performance indicator, in Figure~\ref{fig:7} the values of the Area Under Curve (AUC) parameter~\cite{ROC88} associated with the considered algorithms and simulators are also shown---the higher the AUC, the better the performance.

Optimum values of decision thresholds can be defined considering the point of the ROC curve with minimum Euclidean distance to the point~$(0,1)$ which describes the ideal detector.
Considering optimum threshold values for both algorithms, sensitivity, specificity and DOR for the video streams obtained above are shown in Tables~\ref{tab:3} and~\ref{tab:4} for the software- and hardware-based simulators, respectively.
These results show that the sensitivity is high for both algorithms---MMAD and STRE can effectively identify patients suffering from apneas.
However, the specificity is higher with STRE than with \mbox{MMAD---STRE} can identify patients breathing normally more efficiently than MMAD.
The global measure DOR has clearly higher values for STRE, indicating the better overall performance of this algorithm with respect to MMAD.
\begin{table}[t!]
	\renewcommand{\arraystretch}{1.1}
	\small
	\centering
	\begin{tabular}{cccc}
		\hline
		\bfseries Algorithm & \bfseries $\boldsymbol{\alpha} \, [\%]$ & \bfseries $\boldsymbol{\beta} \, [\%]$ & $\boldsymbol{\Delta}$\\
		\hline
		MMAD & $88.8\%$ & $82.9\%$ & $38.4$\\
		STRE & $91.0\%$ & $86.9\%$ & $67.1$\\
		\hline
	\end{tabular}
	\caption{Detection performance on software-based simulator.}
	\label{tab:3}
\end{table}
\begin{table}[t!]
	\renewcommand{\arraystretch}{1.1}
	\small
	\centering
	\begin{tabular}{cccc}
		\hline
		\bfseries Algorithm & \bfseries $\boldsymbol{\alpha} \, [\%]$ & \bfseries $\boldsymbol{\beta} \, [\%]$ & $\boldsymbol{\Delta}$\\
		\hline
		MMAD & $95.1\%$ & $78.7\%$ & $71.7$\\
		STRE & $92.3\%$ & $89.6\%$ & $103.3$\\
		\hline
	\end{tabular}
	\caption{Detection performance on hardware-based simulator.}
	\label{tab:4}
\end{table}

An in-depth comparison between MMAD and STRE is beyond the scope of this paper.
Nonetheless, the presented results highlight the importance of the proposed breathing CTMC statistical model and simulators for performance analysis and optimized design of video-based monitoring systems.

\section{Conclusion}
\label{sec:Conclusion}
In this paper, a CTMC statistical model describing the breathing behavior of a patient, healthy or suffering from breathing disorders, is presented.
The values of the model parameters are estimated from the analysis of vital signs of hospital-monitored patients, in order to realistically describe RR patterns.
The proposed CTMC model is used to implement two simulators, software- and hardware-based, useful to develop and test video processing-based algorithms to monitor the RR and detect possible apnea events.

The statistical model is validated and the simulators are tested with previously developed systems for RR estimation and apnea event detection.
The results show that the presented model provides a reliable and realistic method to simulate breathing patterns and respiratory pauses/apneas.
This statistical model can be strategic to create extended video databases, useful to design and test video processing-based algorithms for automatic breath monitoring.

\subsubsection*{Standard Protocol Approvals, Registrations, and Patient Consents}
In accordance with current practice at our Institution, an informed consent form was signed by a parent of each newborn patient, and the aforementioned document was stored in the patients' hospital chart.
Analysis and use of biomedical signals and video recordings was approved by the Ethical Local Committee.

\bibliographystyle{IEEEtran}
\bibliography{ModSimBreathPattBIB}
\end{document}